\newcommand{\True}{True}

\newcommand{\IncludeAll}{True} 		
\newcommand{\ListofTables}{False}	
\newcommand{\ListofFigures}{False}	
\newcommand{\IncludeCR}{False}		
\newcommand{\IncludeDed}{True}		
\newcommand{\IncludeAck}{True}		
\newcommand{\IncludeApp}{False}		

\newcommand{\Info}{Info}		
\newcommand{\Aux}{Aux}			
\newcommand{\Chapters}{Chapters}	
\newcommand{\Appendixes}{Appendixes}	

\newread\InStream
\newcount\Done
\def\LOOP#1\REPEAT{\def\BODY{#1}\ITERATE}
\def\ITERATE{\BODY\let\next=\ITERATE\else\let\next=\relax\fi\next}
\def\NoName{}

\newcommand{\InputFiles}[2]{
  \openin\InStream=#1/#2
  \Done=0
  \LOOP\ifnum\Done=0
     \ifeof\InStream\global\Done=1
     \else
     \read\InStream to \FName
     \ifx\NoName\FName{}\else
       \ifeof\InStream{}\else
          \input#1/\FName
       \fi
     \fi
  \fi
  \REPEAT
  \closein\InStream
}

\newcommand{\ReadThesisTitle}{
  \begin{centering}
  \openin\InStream=\Info/Title
  \Done=0
  \LOOP\ifnum\Done=0
     \ifeof\InStream\global\Done=1
     \else
     \read\InStream to \TextLine
     \ifx\NoName\TextLine{}\else
       \ifeof\InStream{}\else
          \TextLine \\
       \fi
     \fi
  \fi
  \REPEAT
  \closein\InStream
  \end{centering}
}

\openin\InStream=\Info/GradInfo
\read\InStream to \FullName
\read\InStream to \ThesisType
\read\InStream to \DegreeType
\read\InStream to \GradMonth
\read\InStream to \GradYear
\read\InStream to \Department
\read\InStream to \Advisor
\closein\InStream

\documentclass[10pt,twoside]{report}
\usepackage{amssymb}
\usepackage{amsmath}
\usepackage{epsfig}
\usepackage[round]{natbib}
\usepackage{\Aux/Thesis}
\begin{document}
\pagestyle{plain}
\pagenumbering{roman}
%

%
\ifx\IncludeAll\True
\SingleSpace
\pagestyle{empty}
\vspace*{33pt}
\SingleSpace
{\bf \Large \ReadThesisTitle}
\vspace*{5.5 cm}
\centerline{By}
\vspace*{.5 cm}
\centerline{\bf \Large \expandafter{\FullName}}
\vspace{7.0 cm}
\centerline{A \uppercase\expandafter{\ThesisType}\ PRESENTED TO THE 
            GRADUATE SCHOOL}
\centerline{OF THE UNIVERSITY OF FLORIDA IN PARTIAL FULFILLMENT}
\centerline{OF THE REQUIREMENTS FOR THE DEGREE OF}
\centerline{\uppercase\expandafter{\DegreeType}}
\vspace{1 cm}
\centerline{UNIVERSITY OF FLORIDA}
\vspace{1 cm}
\centerline{\GradYear}

\cleardoublepage
\fi
%

%
\ifx\IncludeAll\True
  \ifx\IncludeCR\True
    \thispagestyle{empty}
\vspace*{3.5in}
\centerline{Copyright \GradYear}
\vspace*{0.5in}
\centerline{by}
\vspace*{0.5in}
\centerline{\FullName}

    \pagebreak
  \fi
\fi
%

%
\ifx\IncludeAll\True
  \ifx\IncludeDed\True
    \thispagestyle{empty}
\vspace*{63pt}

\DoubleSpace

\begin{center}

{\it \large Dedicated with love to my parents, Vassili and Maria, and to
my sister Eleni \\ \vspace{5mm} for all their support, encouragement and
understanding}

\end{center}

\SingleSpace

    \cleardoublepage
  \fi
\fi
%

%
\pagestyle{plain}
\ifx\IncludeAll\True
  \ifx\IncludeAck\True
    \addcontentsline{toc}{chapter}{ACKNOWLEDGEMENTS}
    \vspace*{63pt}
\centerline{ACKNOWLEDGEMENTS}
\vspace{.5in}

\DoubleSpace

It takes a certain amount of perseverance, love for one's research
subject, and, sometimes, some dose of insanity, to pursue a doctoral
degree. This highest of scholar degrees with which mankind awards
individuals, is usually the culmination and convergence point of most of
the candidate's previous plans and goals in life, and often takes a
lifetime of preparation to achieve. For this reason, I would like to thank
all the wonderful people who helped nurture my interest in science, and
learning in general, during my pre-college years, and especially my
parents, to whom this dissertation is dedicated. They always encouraged me
to follow my dreams and never attempted to interfere except to help me
achieve them, even if that meant that they would have to sacrifice or
alter some of their plans and expectations, and for that I am grateful.

I would like to express my gratitude to my college professors at the
University of Ioannina, Greece, and especially Drs.\ I. E. Lagaris, E.\
Manesis and V.\ Tsikoudi, for showing genuine interest and helping me
achieve my goal of pursuing a graduate degree. I am particularly indebted
to Dr.\ Tsikoudi, my undergraduate professor of Astronomy, who encouraged
my aspirations for a graduate degree throughout my college years, then
spent endless hours working with me, helping me make crucial decisions
during the hectic months before graduate application deadlines, and
supported me and my family when the time had come to leave home and change
continent. It is no exaggeration to say that most probably this
dissertation would not have even started without her personal involvement
and support, and I would like to let her know that they are truly
appreciated.

It would be a grave omission not to express my gratitude to our family
friends, Leonidas Gkelios, a former Air Force and civil aviation pilot,
and his wife, Agni. Their travel experience, tips and advice, but most of
all their moral support to me and my parents during the days before my
departure for the United States and the first weeks thereafter, were
invaluable and will be forever remembered and appreciated.

It is hard for me to imagine that I could have been more fortunate than
having Henry Kandrup as my advisor. Henry masters an impressive
combination of breadth and depth of knowledge and experience that make him
an invaluable resource for a graduate student. At the same time, he shows
a deep respect and personal interest for his students, being always
available, never ``too busy,'' ready to provide his insight whenever
needed, but in such a way that the student will come to realize, on his or
her own, where the error or the problem was, and learn from it. He works
closely with his students to establish what they {\em really} want to do
---occasionally altering his own research plans to accommodate their
interests--- and making it clear all along, in a nonverbal but very
obvious way, that it is the student, and not some narrowly defined
project, what matters most. This experience has been quite overwhelming,
and I can only hope that one day I can show my true appreciation and
gratitude by imitating just a few of these virtues.

It is a great pleasure to acknowledge the close collaboration, for parts
of this dissertation, with Dr. E. Athanassoula of the Observatoire de
Marseille, France. I am indebted for her hospitality and financial support
during my two-month stay at the observatory, but also for the lengthy
discussions and constructive criticisms that have strengthened the
credibility of the results reported in this work. Access to her group's
GRAPE computer facility is instrumental for the ongoing structural
stability studies, briefly described in Chapter~4, but also for pointing
out the necessity of studying the simpler case of a Plummer sphere, in
Chapter~2.

Professor George Contopoulos has certainly been along with me during this
scientific journey for longer than anyone else. His astronomy lecture
notes, from the time when he was at the University of Thessaloniki,
Greece, helped nourish my fledgling interest in astronomy (even
unknowingly to him) during my high-school years, and the lectures and
discussions he has had with me years later, as a visiting professor at the
University of Florida, helped better shape the goals of this research. But
I am also thankful to him and to the European Community Human Capital and
Mobility program for offering me the opportunity to spend part of a summer
at the Institut f\"ur Astronomie of the University of Vienna, Austria, in
collaboration with Dr. Rudolf Dvorak, working on a project that would give
me a better understanding of short-time Lyapunov exponents, a tool that
proved particularly useful in parts of this dissertation. Very special
thanks should go here, of course, to a wonderful Rudi Dvorak, for all the
discussions, the useful comments and the encouragement that he offered,
and for his outstanding hospitality.

At the heart of this research project lies the quadratic programming
method that solves the optimization problem and computes the models on
which most of the conclusions of this work are based. I am truly indebted
to Dr.\ Cs. M{\'e}sz{\'a}ros of the MTA SZTAKI Computer and Automation
Research Institute of the Hungarian Academy of Sciences, who generously
made available, free of charge, recent versions of his quadratic/linear
programming solver, BPMPD, and spent his time answering my questions.
BPMPD is a fine piece of software, robust and very efficient in its use of
computational resources. Without BPMPD, the scope and credibility of this
work would be severely restricted.

Brent Nelson, our computer systems administrator, was at the receiving end
of much of the heat related to this project and he deserves special
thanks. If it were not for his far-sighted strategy of installing a
cost-effective, powerful and versatile cluster of Linux workstations, the
quality of this work would have been severely compromised.

I would also like to thank all of my committee members for their interest
and support, and especially Drs. Haywood Smith, for his careful reading of
the manuscript and his suggestions; Jim Ipser, for the time that he spent
for me on a number of occasions and for his advice; and Steve Gottesman
and his wife Mariou for their interest and encouragement, on academic as
well as culinary and other matters. It is also a pleasure to thank Dr.\
Robert Wilson for his hospitality on a number of occasions, and Dr.\
Heinrich Eichhorn for several discussions on issues related to my
research, but also on a wide range of other topics.

During the years of this work, I have been the beneficiary of advice and
encouragement from a number of people who have done work in this field and
on whose work my research was largely based. I would like to thank Dr.\
David Merritt for providing much of the motivation for doing this work,
and for useful discussions and advice on a number of implementation and
conceptual issues that he kindly offered me on a number of occasions; and
Drs.\ Chris Hunter, Tim de Zeeuw and Ron Buta for encouragement and
helpful advice. 

It is also a great pleasure to acknowledge the help and friendship of a
number of former graduate students (and now doctors) in my department.
First, I would like to thank Dave Kaufmann for many lively discussions,
especially on the Contopoulos-Grosb{\o}l method for constructing
self-consistent models of galaxies ---a great help during my first steps
in this project from a veteran galaxy builder; Seppo Laine for being so
obligingly helpful, especially during my first year of adaptation to life
in Gainesville, and for all the adventures that we have been to
thereafter; Jaydeep Mukherjee, for being such a great friend, counselor
and mentor; and Dimitri Pourbaix, for his obliging assistance with a
number of numerical analysis and computer-related problems, as well as for
the endless pleasure of debating with him on any and all issues.

Thanks to the vigilance of our three office ladies, Ann Elton, Glenda
Smith and Debra Hunter, many administrative pitholes were averted and
others were promptly remedied. Their assistance, care and smiling faces
are appreciated.

The financial and moral support of the Department of Astronomy is also
gratefully appreciated. Furthermore, I would like to acknowledge a
Fulbright Fellowship that greatly facilitated my first year of graduate
studies, a Research Assistantship from the Florida Space Grant Consortium
(NASA), and a Sigma-Xi Grant-in-Aid of Research which was essential in
allowing me to travel to Marseille.

Finally, I would like to thank the people of Florida and of the United
States for their hospitality and accommodation during my years of study at
the University of Florida.

\SingleSpace

    \newpage
    \pagestyle{empty}
    \cleardoublepage
  \fi
\fi
%

%
\pagestyle{plain}
\SingleSpace
\tableofcontents
\cleardoublepage
%

%
\ifx\ListofTables\True
  \DoubleSpace
  \addcontentsline{toc}{chapter}{LIST OF TABLES}
  \listoftables
  \pagebreak
\fi
%

%
\ifx\ListofFigures\True
  \addcontentsline{toc}{chapter}{LIST OF FIGURES}
  \listoffigures
  \pagebreak
\fi
%

%
\ifx\IncludeAll\True
\addcontentsline{toc}{chapter}{ABSTRACT}
\SingleSpace

\centerline{Abstract of \ThesisType\ Presented to the Graduate School}
\centerline{of the University of Florida in Partial Fulfillment of the}
\centerline{Requirements for the Degree of \DegreeType}
\vspace*{2 cm}
\ReadThesisTitle
\vspace*{.5 cm}
\centerline{By}
\vspace*{.5 cm}
\centerline{\FullName}
\vspace*{.5 cm}
\centerline{\GradMonth\ \GradYear}
\vspace*{1 cm}
\centerline{Chairman: Dr. \Advisor \hspace*{\fill}}
\centerline{Major Department: \Department \hspace*{\fill}}
\vspace*{5 mm}

\DoubleSpace


Schwarzschild's method was used to construct equilibrium solutions to the
collisionless Boltzmann equation corresponding to a Plummer sphere. These
solutions were compared with analytical results to test the robustness of
the numerical method and its efficiency in probing the degeneracy of the
solution space. The method was then used to construct genuinely triaxial
stellar equilibria, for which no analytical solutions are known, and to
study their nonuniqueness. The particular model that was studied is a
three-dimensional generalization of Dehnen's spherical potential, which
contains a central density cusp and admits both regular and chaotic
orbits. It was found that, for a model with a weak density cusp,
self-consistent models do not exist if the chaotic orbits are assumed to
be completely mixed so as to yield a time-independent building-block: only
the innermost 65\% of the mass can be mixed. In these inner regions, it is
possible to obtain alternative solutions that contain significantly
different numbers of chaotic orbits, yet yield (at least approximately)
the same mass density distribution. However, it is not clear whether these
solutions are truly time-independent, since the ``unmixed'' chaotic orbits
in the outer regions, which do not sample an invariant measure, can cause
secular evolution. When using Schwarzschild's method, one must be very
careful to sample accessible phase-space as comprehensively and as densely
as possible, while at the same time ensuring that each orbit is a truly
time-independent building block. Some of the numerical equilibria were
sampled to generate initial conditions for $N$-body simulations, with the
aim of testing the structural stability of the models. Preliminary work
showed that no catastrophic evolution takes place, but there is a weak
tendency for the configuration to become more nearly axisymmetric over a
period of several dynamical times. It is not presently clear whether this
tendency is real or a numerical artifact.

\SingleSpace

\newpage
\pagestyle{empty}
\cleardoublepage
\fi
%

%
\addcontentsline{toc}{startchapter}{\ShiftHack CHAPTERS}
\DoubleSpace
\pagenumbering{arabic}
\pagestyle{myheadings}
\InputFiles{\Chapters}{ChapterList}
%

%
\ifx\IncludeApp\True

  \DoubleSpace
  \addcontentsline{toc}{startchapter}{\ShiftHack APPENDIXES}
  \appendix

  \InputFiles{\Appendixes}{AppendixList}

\fi
%

%
\newpage
\pagestyle{empty}
\cleardoublepage
\pagestyle{myheadings}
\SingleSpace
\addcontentsline{toc}{chapter}{REFERENCES}
\bibliographystyle{my_plainnat}
\bibliography{Thesis}
%

%
%

%
%

\end{document}